# Artificial Intelligence and Cost Reduction in Public Higher Education: A Scoping Review of Emerging Evidence

Diamanto Tzanoulinou[†], Loukas Triantafyllopoulos[†*],
George Vorvilas[‡], Evgenia Paxinou[†], Nikolaos Karousos[†],
Thomas Dasaklis[§], Athanassios Mihiotis[§], Manolis Koutouzis[‡],
Dimitris Kalles[†], Vassilios S. Verykios[†]

**Abstract**

Public higher education systems face increasing financial pressures from expanding student populations, rising operational costs, and persistent demands for equitable access. Artificial Intelligence (AI), including generative tools such as ChatGPT, learning analytics, intelligent tutoring systems, and predictive models, has been proposed as a means of enhancing efficiency and reducing costs. This study conducts a scoping review of the literature on AI applications in public higher education, based on systematic searches in Scopus and IEEE Xplore that identified 241 records, of which 21 empirical studies met predefined eligibility criteria and were thematically analyzed. The findings show that AI enables cost savings by automating administrative tasks, optimizing resource allocation, supporting personalized learning at scale, and applying predictive analytics to improve student retention and institutional planning. At the same time, concerns emerge regarding implementation costs, unequal access across institutions, and risks of widening digital divides. Overall, the thematic analysis highlights both the promises and limitations of AI-driven cost reduction in higher education, offering insights for policymakers, university administrators, and educators on the economic implications of AI adoption, while also pointing to gaps that warrant further empirical research.

**JEL Classifications:** I22, I23, O33, C45, C55

**Keywords:** Artificial Intelligence; Public higher education; Cost Reduction; Resource Allocation; Affordability

## 1   Introduction

Public higher education is under increasing financial pressure globally, driven by rising enrolments, escalating operational costs, and persistent demands for more equitable access (Al-Zahrani and Alasmari, 2024; Wang, 2024). These pressures include costs for infrastructure, staffing, administration, and student support services, all compounded by technological advancement and the expectations of digital readiness (Li et al., 2021). The mismatch between budgets and rising costs is especially acute for public institutions with limited funding, often leading to trade-offs that can reduce quality, limit access, or delay innovation (Al-Zahrani and Alasmari, 2024).

---

[†] School of Science and Technology, Hellenic Open University, Patras, Greece
[*] Corresponding author. School of Science and Technology, Hellenic Open University, Patras, Greece.          Email: triantafillopoulos.loukas@eap.gr
[‡] School of Humanities, Hellenic Open University, Patras, Greece
[§] School of Social Sciences, Hellenic Open University, Patras, Greece

Artificial Intelligence (AI) is increasingly viewed as a mechanism to help address these challenges by improving efficiency, automating routine tasks, and enabling more scalable educational models (Goel, 2020; Wang, 2024). For example, AI-powered virtual assistants and adaptive learning systems have been shown to reduce instructor workload while maintaining or improving learning outcomes (Goel, 2020). Generative AI, learning analytics, and predictive models also promise to help institutions anticipate student needs (e.g., risk of dropout) and optimize resource allocation (McConvey et al., 2023).

Despite these promising developments, relatively few studies have examined AI specifically through the lens of cost reduction and economic efficiency in public higher education. Many works focus on pedagogical outcomes, student experience, or ethical and social implications (Wang, 2024). Less is known about where costs are saved, what the upfront investments are, how sustainable the savings are, or how these technologies perform in less-resourced settings. Also, ethical, governance, and equity issues are often mentioned but not deeply analyzed in terms of their cost implications (McConvey et al., 2023).

The present study therefore conducts a scoping review to map out the emerging evidence on how AI contributes to cost reduction, efficiency, and affordability in public higher education. Key research questions include:

(1) In which areas or functions do public universities use AI to reduce costs?
(2) What are the benefits and limitations observed in those cases?
(3) What institutional, infrastructural, or contextual factors influence the success or failure of AI for economic efficiency?
(4) What gaps remain in the literature that require further empirical study?

Guided by these questions, the paper first presents the scoping-review methodology (Section 2), then addresses the questions via a thematic synthesis of five AI application domains (Section 3), before turning to a critical appraisal of opportunities, constraints, and limitations (Section 4), and concluding with implications for policy, institutional decision-making, and future research (Section 5).

## 2  Materials and Methods

This section describes the methods employed to locate and choose studies for inclusion, covering the formulation of eligibility criteria, the search approach, and the screening procedure.

## 2.1 Study Design

The review followed the PRISMA 2020 guidelines (Page et al., 2021), which offer a standardized framework for reporting systematic reviews.

## 2.2 Eligibility Criteria

Clear eligibility criteria are critical to maintaining the validity, relevance, and focus of any systematic review (Sohrabi et al., 2021). In this study, exclusion rules were applied in a sequential order: once an article failed to meet the first condition, it was not assessed against the remaining ones. For inclusion, a study was required to satisfy all of the following criteria:

- IC-1: The study applies AI methods (e.g., machine learning, deep learning, generative AI such as ChatGPT, learning analytics, intelligent tutoring systems, or AI-driven decision support).
- IC-2: The context is public higher education institutions (HEIs), or the findings are directly transferable to public HE.
- IC-3: The study provides empirical evidence on economic/efficiency-related outcomes (e.g., cost reduction, cost-effectiveness, affordability, expenditure, resource allocation).
- IC-4: The study is a peer-reviewed journal article or conference proceeding.
- IC-5: The full text is available in English.

In contrast, studies were excluded if they met any of the following conditions, which were assessed in sequential order:

- EC-1: The study does not apply AI methods (e.g., ICT, digital tools, or simulation only, without AI).
- EC-2: The study does not involve public higher education (e.g., K-12, vocational training, private corporate training, healthcare, or industry).
- EC-3: The study does not provide empirical evidence on cost, efficiency, affordability, or resource allocation outcomes.
- EC-4: The publication is not peer-reviewed (e.g., book chapter, white paper, editorial) or the full text is inaccessible.
- EC-5: The article is written in a language other than English.

## 2.3 Search Strategy

The search strategy was developed to systematically identify peer-reviewed research at the intersection of three thematic domains:

1. AI methods and applications;
2. Public higher education institutions (HEIs);
3. Cost, efficiency, and resource allocation outcomes.

To ensure comprehensive and reliable coverage, both the Scopus and IEEE Xplore databases were selected as primary sources, given their strong representation of high-impact research across education, computer science, and economics. Using these two databases ensures consistency and replicability in retrieval.

The search was conducted on September 4, 2025, and employed a structured combination of keyword groups. Specifically:

- AI-related terms included: "generative AI", "ChatGPT", "artificial intelligence", "machine learning", "learning analytics" and "intelligent tutoring systems".
- Higher education terms included: "public higher education" and "public university/ies".
- Cost and efficiency terms included: "cost reduction", "cost-effectiveness", "affordability", "educational expenditure", and "resource allocation".

The query was designed to be compatible with both Scopus and IEEE Xplore advanced search syntaxes in order to maximize recall and precision. The final search string is presented in Table 1.

**Table 1: Search strings used across databases**

| Database | Search Query |
|---|---|
| Scopus | ("generative AI" OR "ChatGPT" OR "artificial intelligence" OR "machine learning" OR "deep learning" OR "learning analytics" OR "intelligent tutoring systems") AND ("public higher education" OR "public university" OR "public universities") AND ("cost reduction" OR "cost-effectiveness" OR "affordability" OR "educational expenditure" OR "resource allocation") |
| IEEE Xplore | (("All Metadata":"generative AI" OR "All Metadata":"ChatGPT" OR "All Metadata":"artificial intelligence" OR "All Metadata":"machine learning" OR "All Metadata":"deep learning" OR "All Metadata":"learning analytics" OR "All Metadata":"intelligent tutoring systems") AND ("All Metadata":"public higher education" OR "All Metadata":"public university" OR "All Metadata":"public universities") AND ("All Metadata":"cost reduction" OR "All Metadata":"cost-effectiveness" OR "All Metadata":"affordability" OR "All Metadata":"educational expenditure" OR "All Metadata":"resource allocation")) |

### 2.4 Study Selection Process

Two reviewers independently carried out the screening process, beginning with titles and abstracts to determine initial relevance. Subsequently, full texts were examined against the eligibility criteria described in Section 2.2. Any disagreements about whether to include or exclude a study were resolved through discussion, ensuring alignment with the review's scope.

### 2.5 Study Selection Process

After applying the PRISMA 2020 screening process, 21 studies were retained for systematic review and data extraction. From each study, we recorded essential details including:

- The year of publication;

- The institutional context;

- The type of AI application (e.g., generative AI, learning analytics, intelligent tutoring systems, decision support);

- The reported outcomes related to cost reduction or efficiency.

To facilitate comparison and synthesis, the studies were organized thematically by both focus and methodological approach. The resulting categories — student advising and retention, teaching and learning support, institutional resource management, productivity and policy evaluation, and equity considerations — capture the key domains of AI use in public higher education and form the basis for Section 3, which examines how AI technologies are being leveraged to enhance efficiency and affordability in the sector.

### 2.6 Risk of Bias

Despite employing systematic procedures, including independent dual screening and consensus-based inclusion decisions, certain risks of bias may remain. Language bias is possible, as only English-language publications were included (EC-5). Database bias may also be present, since the search was limited to Scopus and IEEE Xplore; however, the IEEE search yielded only a single article, which was already indexed in Scopus, so relevant studies indexed in other databases or in grey literature may still have been missed. In addition, the evolving terminology of AI could have introduced bias, with some relevant works overlooked if alternative terms were not captured by the search strategy. Finally, interpretive bias may have occurred during data extraction and categorization, particularly where cost- or efficiency-related outcomes were implicit rather than explicitly reported. While these risks were mitigated through predefined inclusion/exclusion criteria, transparent reporting, and reviewer discussions, they cannot be entirely ruled out given the exploratory scope of a scoping review.

## 3 Results

The next section presents the findings of the scoping review, beginning with the identification and selection of eligible studies. This is followed by a descriptive summary of the included literature and a thematic analysis of the main domains in which AI has been applied to enhance efficiency and reduce costs in public higher education.

### 3.1 Identification and Inclusion of Studies

The search strategy identified a total of 241 records, with 240 retrieved from Scopus and 1 from IEEE Xplore. No additional records were obtained from registers or other sources. One duplicate record was removed prior to screening, leaving 240 records for title and abstract screening. Of these, 210 were excluded as not meeting the eligibility criteria. The remaining 30 reports were sought for retrieval, but 9 could not be accessed in full text. The 21 reports retrieved were assessed for eligibility, all of which satisfied the predefined inclusion criteria and were included in the final review. The complete selection process is summarized in the PRISMA 2020 flow diagram (Figure 1), which was generated using the PRISMA2020 Shiny app and R package developed by Haddaway et al. (2022).

### 3.2 Descriptive Overview of Included Studies

The 21 studies included in the review were analyzed to provide a descriptive overview of recent research trends on AI-driven cost reduction strategies in public higher education. In particular, we examined their distribution by year of publication and publisher patterns.

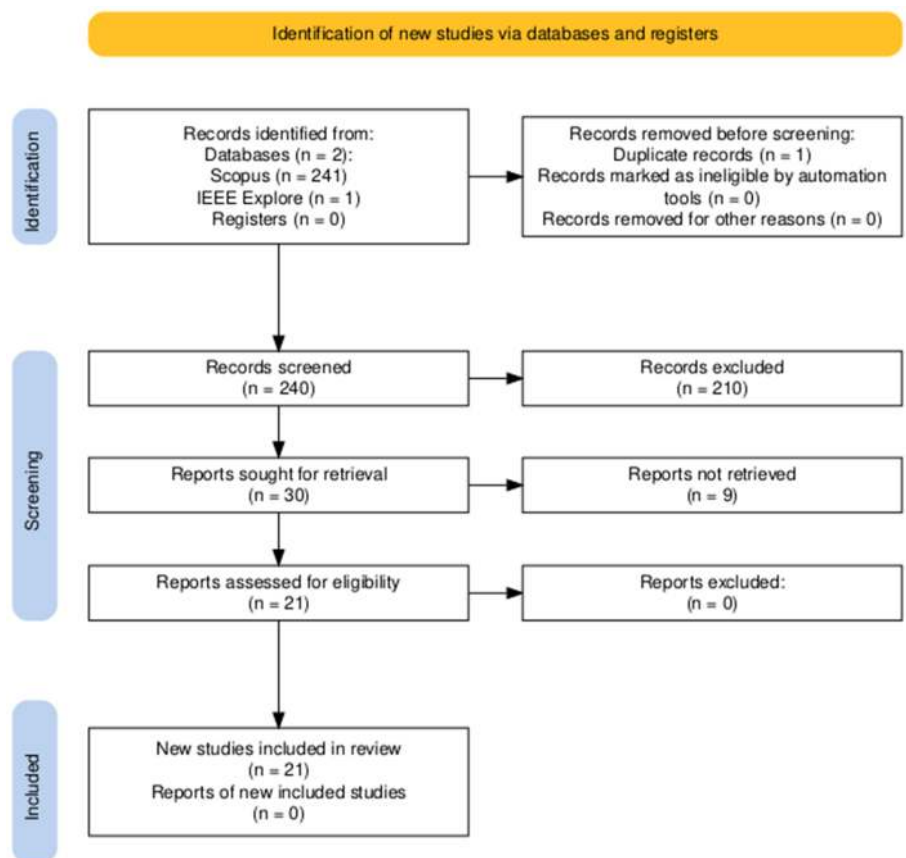

**Figure 1. PRISMA 2020 flow diagram.**

As shown in Figure 2, the temporal distribution of publications reveals an early contribution in 2005, followed by a resurgence beginning in 2019. Research output remained steady in 2019–2020, dipped in 2021, and then grew again from 2022 onward. The sharpest increases occurred in 2023 and 2024, with 2024 marking the peak of publication activity. Although 2025 shows a slight decline from this peak, the number of studies remains higher than in all years prior to 2023. This trend reflects the accelerating academic interest in AI — particularly generative applications, learning analytics, and intelligent tutoring systems — as tools for enhancing efficiency and reducing costs in public higher education.

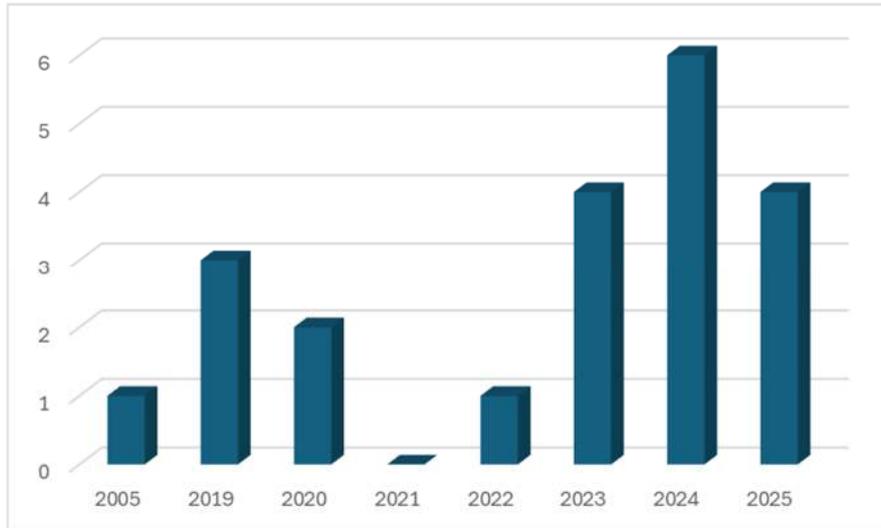

**Figure 2. Temporal distribution of publications on AI-driven cost reduction strategies in public higher education**

Furthermore, regarding dissemination venues, Figure 3 shows that research on AI-driven cost reduction in public higher education has appeared across a diverse set of journals, highlighting its cross-disciplinary nature. The journals Lecture Notes in Computer Science, Applied Mathematics and Nonlinear Sciences, and Sustainability (Switzerland) each account for two publications, representing the largest shares. The remaining studies are distributed across a wide range of outlets, including education-focused journals (American Journal of Distance Education, Higher Learning Research Communications, Journal of Educational Data Mining, International Journal of Web-Based Learning and Teaching Technologies, Marketing Education Review), technology and computer science venues (IEEE Access, IAES International Journal of Artificial Intelligence, International Journal of Human-Computer Interaction), health-related and multidisciplinary platforms (Health Science Reports, International Journal of Medical Informatics, Discover Artificial Intelligence, Benchmarking, Journal of Applied Research and Technology).

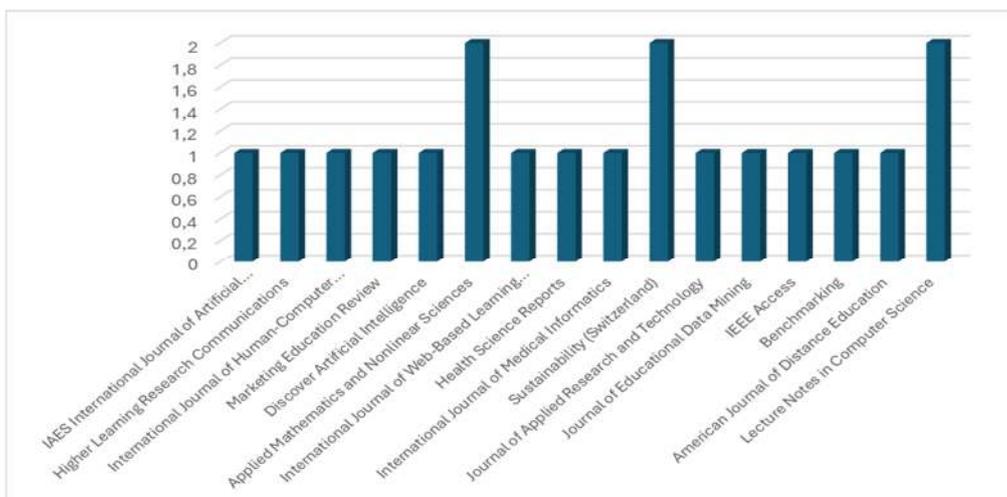

**Figure 3. Publication venues of studies on AI-driven cost reduction in public higher education**

### 3.3   Thematic Analysis of AI Applications for Efficiency and Affordability in HEIs

To understand how AI is used to improve efficiency and reduce costs in public higher education, we conducted an inductive thematic analysis of the included studies and grouped them into five AI-focused domains: (1) AI-enabled resource optimization and timetabling; (2) AI-driven affordability and access via financial optimization (e.g., scholarship/aid); (3) AI-powered digitalization of teaching and learning infrastructure; (4) AI-supported administrative automation and institutional resilience; and (5) AI-mediated skills development and lower-cost training. These domains reflect recurrent challenges for public universities — rising operating costs and growing enrollments — as well as the computational approaches deployed to address them. The resulting taxonomy is grounded in recurring patterns across the literature and guides the synthesis that follows.

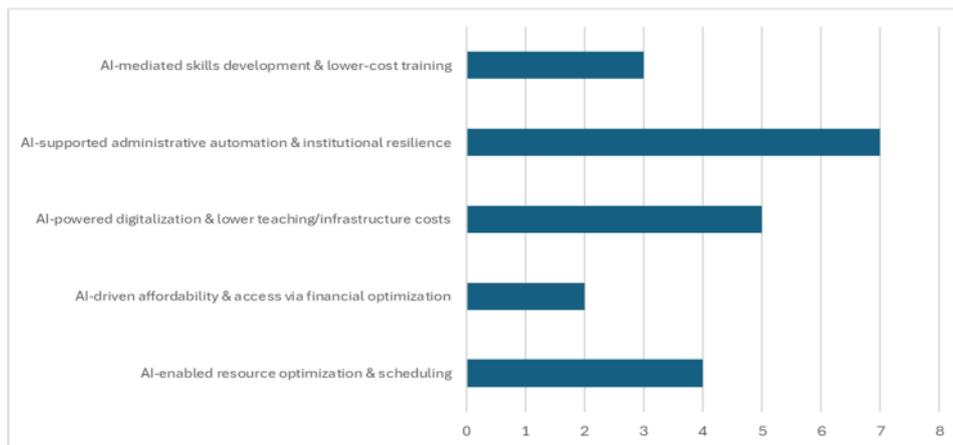

**Figure 4: Key Areas of AI Application for Cost Reduction in public higher education**

As shown in Figure 4, the most common domain is AI-supported administrative automation and institutional resilience (n=7), followed by AI-powered digitalization and lower teaching/infrastructure costs (n=5) and AI-enabled resource optimization and scheduling (n=4). The remaining areas are AI-mediated skills development and lower-cost training (n=3) and AI-driven affordability and access via financial optimization (n=2). Taken together, this pattern highlights a primary emphasis on administrative and infrastructural efficiencies, with parallel attention to affordability and workforce development. Each of the following subsections examines these five domains in turn, outlining the methods used and synthesizing the main findings.

### 3.3.1 AI-mediated skills development and lower-cost training

In the first thematic area, three empirical studies were identified that examine how AI can mediate skills development while lowering the marginal cost of training delivery — primarily by replacing or augmenting labor-intensive workshops with scalable, self-paced systems and by directing instructional time where it yields the greatest return.

Ugwu et al. (2025) report a randomized pretest–posttest trial with 124 non-English early-career researchers at a Nigerian public university, comparing eight weeks of AI-based immersive, self-directed English training with equally timed human-led workshops. Their repeated-measures ANOVA shows a large group effect [$F(1, 122) = 38.12$, $p<.001$, $\eta p^2=0.24$] and a significant time×group interaction [$F_{(2, 244)} = 31.87$, $p<.001$, $\eta p^2=0.21$], with the AI group outperforming the workshop group at posttest and three-month follow-up; Tukey post hoc tests confirmed sustained gains for the AI condition. These findings, obtained under matched "contact hours" and controlled exposure, indicate at least comparable, and often superior, learning outcomes when staff time is substituted by an adaptive platform, implying lower unit costs at scale because additional learners do not require proportional increases in instructor labor. The trial design and dosage matching are detailed in their methods (random allocation; 64 hours over eight weeks in both arms), strengthening the cost-relevance of the substitution logic. The magnitude of gains is documented in the estimated marginal means and post-hoc tables.

Turning to generative AI in routine study support, Jaboob et al. (2025) analyze survey data from 768 students across three Arab countries using PLS-SEM. They find that both generative-AI techniques and applications have significant positive effects on student behavior and cognitive achievement (e.g., direct path to achievement Original Sample=0.344; T-Statistics=9.714; $p<.001$), with behavior partially mediating the relationship (indirect effects H6 and H7 significant). While the study does not compute costs, the mechanism—improved achievement and study behaviors via on-demand AI tools — suggests a route to "do more with the same": universities can lift outcomes without commensurate increases in tutoring staff or contact hours, thereby spreading fixed support costs over better results.

A complementary strand focuses on sensing and targeting engagement so that scarce instructional time is spent where it matters most. Wang (2023) develops an information-fusion framework that uses face localization (Haar features), CNN-based expression analysis, eye-state (PERCLOS) fatigue detection, and face-pose estimation to derive a composite classroom engagement score; decision-level fusion improves classification performance (e.g., F1≈79% for expression features). By fusing emotion, concentration, fatigue, and answer-accuracy signals, the system can triage support to learners who are disengaging, helping instructors target remediation instead of delivering uniformly intensive support to all students—an efficiency gain with clear cost implications, even though the paper itself focuses on technical performance rather than economics.

### 3.3.2 AI-supported administrative automation and institutional resilience

The most common thematic area — AI-supported administrative automation and institutional resilience — includes seven empirical studies. Collectively, these show how NLP/text-mining, process mining, and machine-learning pipelines can replace slow, staff-intensive fieldwork and manual processing, while providing managers with near-real-time evidence to target funds, policies, and communications more efficiently.

An early contribution comes from Biazus et al. (2019), who applied process mining within a business process management (BPM) approach to analyze and optimize administrative workflows in public universities. By mapping event logs and identifying inefficiencies in procedures such as enrollment and records management, their approach highlighted bottlenecks, rework, and delays that inflate operating costs. The study demonstrated how data-driven workflow analytics can guide redesign efforts, leading to shorter processing times, reduced staff workload, and more consistent service quality — suggesting potential links to lower administrative expenditure.

Abdul-Rahman et al. (2023b) used programmatic mining of Twitter, topic modeling (LDA), and sentiment analysis (VADER) to surface and rank local challenges across six towns on six continents; the outputs are then validated through an online expert survey. This mixed AI + expert design turns unstructured user-genera ted content into decision-ready evidence — "simple, fast, cheap, and efficient" compared with conventional surveys — so planners can prioritize problems with minimal staff time and travel (e.g., noise, housing pressure, governance frictions). Methodologically, the pipeline mines and filters tweets per detected topic and computes negative–polarity frequencies as a proxy for community stressors before expert ranking; the paper details the LDA/VADER steps and the validation workflow. The cost logic is direct: automated, remote sensing replaces repeated in-person consultations for initial scoping, enabling cheaper iteration and faster feedback cycles.

Building on that foundation, a second 2023 study develops a Composite Resilience Index (CRI) localized to Akoka (Lagos) by combining AI-prioritized social media signals with Delphi and analytic hierarchy process (AHP) for weighting (Abdul-Rahman et al., 2023b). Here the data backbone is unusually large — 935,822 tweets (2010–2021) mined from residents and visitors — used to derive candidate indicators before expert elicitation and weighting; the resulting CRI concentrates ≥70% of total weight on physical, economic, social, and cultural criteria, guiding where limited resources can do the best. As the authors note, this approach is a "resource-efficient" alternative to small-sample household surveys, yielding a replicable index that public universities and municipalities can update without extensive field deployments — lowering recurrent monitoring costs while improving targeting precision.

Extending this focus on data-driven resilience and operational efficiency in Brazil, Abonizio et al. (2023) present CoronaAI, a WhatsApp chatbot developed at a public university that uses NLP/Dialogflow to answer health queries and counter misinformation. Analyzing ~7,000 interactions over eleven months, the team shows how AI-mediated, automated response workflows can handle routine information requests at scale, reducing staff time on repetitive inquiries while maintaining service continuity during demand spikes. As an always-on, remotely updatable channel, CoronaAI illustrates how AI-supported administrative automation can bolster institutional resilience, delivering timely guidance without specialized infrastructure or costly field deployments.

Sani and Mandina (2024) focus on digital HRM in the public sector — a setting that includes public universities — mapping the barriers and enablers of adopting AI-supported HR processes (e.g., automated applicant screening, leave/payroll processing, workflow routing, and records management). As a policy and practice-oriented book chapter, they synthesize literature and cases to identify capacity constraints (skills, infrastructure), governance and compliance needs, and change-management requirements, then articulate how targeted training and leadership support are aimed at administrative time savings and cycle-time reductions in core HR tasks. The cost logic is direct: once routine HR workflows are digitized/automated, institutions reduce manual handling, rework, and paper-based overhead, improving resilience by sustaining service levels under tight budgets.

Turning to the public discourse that shapes policy, Göçen et al. (2024) apply sentiment analysis and topic modeling to 157,943 English-language tweets (January–May 2023), a period spanning the surge of interest in tools like ChatGPT. Their time series reveals large-scale dynamics and finance-linked themes ("cost," "debt," "funding," "loan," "expensive") alongside access/equity keywords; overall sentiment about higher education skews positive, while AI mentions tend more negative. For universities, this functions as a low-cost, always-on policy radar: automated NLP surfaces affordability concerns and acceptance risks early, allowing adjustments to aid communications, outreach, or service design before small issues become expensive crises.

Recently, Nai et al. (2025) extended the same logic of administrative automation and institutional resilience to the system level by profiling and forecasting Moroccan education indicators with machine-learning models applied to World Bank data (253 indicators, focusing on dropout, teacher training, expenditure, support-program coverage, and international assessment performance). In a head-to-head comparison (linear/polynomial regression, decision tree, SVR, MLP, random forest), random forest performs best (MAE=3.14; RMSE=3.91; $R^2$=0.92), enabling more reliable projections that directly inform budgeting and staffing decisions. By sharpening forecast precision for dropout and support indicators, this pipeline operationalizes data-driven administration—reallocating funds toward higher-yield interventions and avoiding overspend from misforecasted demand—thereby strengthening resilience at both ministerial and university levels.

### 3.3.3 AI-powered digitalization of teaching and learning infrastructure

This thematic area — AI-powered digitalization of teaching and learning infrastructure — contains a large number of empirical studies in our review (n=5). Together they show how intelligent tutors, algorithmic/IoT classroom platforms, edge-AI architectures, and AI-assisted analytics shift delivery away from fixed, staff- and room-intensive models toward scalable digital workflows and reusable content.

Cung et al. (2019) examined a full-term developmental-mathematics course delivered with the ALEKS intelligent tutoring system, comparing a fully online section to a blended format in a quasi-experimental design while pairing outcomes with an itemized cost model; for a 480-student cohort the authors estimate $53 per student online vs $124 blended, and $142 per passing student online vs $186 blended — evidence that AI-mediated online delivery can reduce per-student expenditure even though academic outcomes were lower online. The study's framing — cost per outcome rather than inputs alone — helps institutions judge when moving substantial instruction into an ITS is financially defensible.

Additionally, Casino et al. (2020) contribute from an infrastructure perspective, enhancing wireless channel estimation through a hybrid ray-launching and collaborative-filtering technique. By filtering noisy simulation cells and reducing collected data size by over 40% (and database patterns by more than 30%), while simultaneously improving accuracy by above 8%, their approach enables universities to model and optimize high-density wireless environments (e.g., campus IoT, smart classrooms) at significantly lower computational and storage costs. This type of AI-assisted optimization ensures that digital teaching platforms can be deployed and scaled with greater efficiency, reducing the expense of maintaining robust connectivity across complex learning spaces.

Furthermore, Tan and Dabbagh (2024) developed an IoT-based interactive teaching platform that applies the Kruskal minimum-spanning-tree algorithm with a Gaussian classification layer to drive adaptive orchestration of activities; the implementation involved 631 university students, with post-course indicators showing higher understanding and engagement for the experimental group than the control. Mid-course analytics and automated orchestration reduce repetitive instructor tasks and enable content reuse at scale — mechanisms that plausibly lower marginal delivery costs even though direct budget figures are not reported.

A complementary contribution by Wang and Zhang (2024) proposes a VR edge-cloud classroom, pairing edge computing with a DIBR-based collaborative transmission and resource-allocation strategy to support multi-user VR. In a controlled comparison (N=50 VR class; N=50 traditional), independent-samples t-tests show significantly higher performance for the VR group across all Bloom levels (including analysis, evaluation, creation). Meanwhile, the edge architecture caches/processes at the network boundary, reducing bandwidth pressure and offloading computing from the cloud, pointing to potential cost efficiencies via shared devices and fewer centralized bottlenecks. These infrastructure choices matter economically when courses scale across sections or cohorts.

Finally, Song et al. (2025) analyzed 322 student responses using computer-assisted semantic network analysis and AI-assisted thematic coding; both traditional and non-traditional students value accessibility/portability, while non-traditional students emphasize affordability. As e-texts typically lower textbook expenditures for students and reduce institutional logistics (printing, warehousing, distribution), the study's design guidance (e.g., offline access, clearer navigation) helps preserve those cost advantages as AI-enhanced features (embedded tutors, adaptive quizzes) proliferate.

### 3.3.4 AI-driven affordability and access via financial optimization

The fourth thematic area is AI-driven affordability and access via financial optimization. Within this area, two empirical studies were recognized for using machine learning coupled with optimization to redesign aid/scholarship policies so that universities can broaden access and meet budget constraints.

Aulck et al. (2020) conducted a large-scale study at a U.S. public university that integrated enrollment-prediction models with a genetic algorithm to optimize the distribution of merit-based scholarships for domestic non-resident applicants. Seven machine-learning classifiers were compared (e.g., XGB, RF, MLP), with the best-performing model providing enrollment probabilities that served as the genetic algorithm's fitness function under explicit fiscal and policy constraints. Using data from 72,589 applicants (2014–2017) and optimizing awards for the 2018 cohort, the approach yielded a 23.3% increase over the historical baseline (actual yield 14.8% vs 12%), compared with a projected 15.8% increase (projected yield 13.9% vs 12%). Beyond revenue growth, the framework reduced reliance on ad hoc consulting and policy revisions, illustrating how AI-enabled financial optimization can deliver both fiscal and operational efficiencies.

Expanding on this logic, Phan et al. (2022) developed a multi-objective optimization framework that simultaneously allocates merit- and need-based aid to balance institutional and student-centered goals — specifically, enrollment, revenue, affordability, and accessibility. The system couples a gradient-boosting classifier (AUC ≈ 0.96) with local-search metaheuristics (stochastic hill climbing and simulated quenching) that dynamically update predicted outcomes as candidate policies evolve. In tests using a recent admissions dataset, the framework increased expected enrollment by 31.9-32,3%, while identifying seven budget-feasible strategies that more than doubled accessibility (≈105–112%) with up to 7% increase in total aid expenditure. By excluding demographic variables to mitigate bias and publishing open-source code, the study demonstrates a portable, low-cost decision-support tool for public HEIs seeking to enhance financial sustainability and equity simultaneously.

### 3.3.5 AI-enabled resource optimization and scheduling

The final thematic area identified in this review concerns AI-enabled resource optimization and scheduling. Within this area, four empirical studies are recognized for demonstrating cost reductions in terms of administrative time, improved utilization of space and staff, and avoided travel. Presented chronologically, they map a shift from early decision-support to simulation-optimization and, more recently, to constraint-based timetabling at institutional scale.

The earliest study (Vinnik and Scholl, 2005) introduced UNICAP, a decision-support system for strategic capacity planning that models educational supply (teaching capacity) and educational demand (curricular consumption). Methodologically, it centralizes heterogeneous institutional data and supports what-if simulations against utilization thresholds before admissions targets are fixed. The empirical application — planning for a new Bioinformatics master's intake — showed how the system flagged an over-capacity condition in one faculty and guided a feasible admissions-target adjustment, helping prevent staff/room misallocations and the costly, reactive corrections that typically follow.

Advancing this focus on institutional efficiency, Kumar (2019) analyzed the performance of India's top 50 management institutes using a hybrid analytics approach — Data Envelopment Analysis (DEA) to compute relative efficiency scores and Kohonen self-organizing maps (an AI/ANN method) to cluster strategies and identify improvement levers. Drawing on data from the National Institutional Ranking Framework (NIRF 2017/2018), the study found significant underutilization among 14 institutions (mean DEA efficiency ≈ 0.68) relative to the sample-wide mean of ≈0.91 across all 50 institutions, implying substantial slack that could be reduced by increasing outputs (e.g., research, consultancy) without adding inputs. This evidence has clear implications for cost containment, showing how AI-assisted benchmarking can guide leaders toward sustainable, low-cost strategies emphasizing productivity and reputation gains over short-term expansion.

Moving from strategic planning to operational deployment, Rios-Esparza and Segura-Pérez (2023) developed a hexagonal tessellation and simulation framework to assign students to host organizations within a public-university social-service program in Mexico City. Their approach jointly optimizes travel distance, scheduling, and skill matching, then stress-tests the assignments through simulation. Applied to a biannual placement cycle, the model reduced total travel distance by 585 km, improved schedule matching by 26% and skills matching by 37% and cut assignment time from 5 days to 90 minutes. These results translate into tangible operational savings in transportation expenditures and administrative labor, achieved without additional infrastructure investments.

Finally, Moreira and Freitas (2024) applied constraint-based optimization to the university timetabling problem using Google's CP-SAT (OR-Tools) solver. Their formulation enforces hard constraints (e.g., instructor availability, avoidance of scheduling overlaps) while minimizing penalties for soft constraints (e.g., preferred times, room consistency) within a single optimization step. Tested on real institutional data, the model systematically reduced constraint violations and manual rework, yielding denser and more stable room utilization and lowering the administrative burden of scheduling. These outcomes demonstrate how AI-enabled scheduling automation can streamline resource allocation and deliver sustained cost reductions for public higher education institutions.

To consolidate these findings and enable cross-study comparison, Table 2 presents a comparative summary of the selected empirical studies on AI in public higher education, detailing the AI approach, cost dimension (e.g., labor substitution, workflow automation, resource optimization), and reported outcomes.

## 4 Discussion

The findings of this scoping review suggest that AI can support cost reduction and efficiency in multiple domains of public higher education. Applications such as intelligent tutoring systems have demonstrated measurable decreases in per-student costs without compromising learning outcomes (Cung et al., 2019), while optimization models for resource allocation and scheduling reduced administrative workload and infrastructure misallocations (Vinnik and Scholl, 2005; Moreira and De Freitas, 2024). Similarly, predictive analytics have been used to improve student retention and enrollment management, thereby reducing financial losses from attrition and misaligned capacity (Aulck et al., 2020). Together, these studies highlight that AI has the potential to enhance institutional sustainability by streamlining both teaching and administrative functions. To convert these efficiency gains into durable benefits, public systems can adopt reinvestment policies that earmark verified AI-related savings for need-based aid, open educational resources, and shared digital infrastructure, thereby stabilizing funding for affordability and capacity building.

At the same time, the analysis underscores several limitations and trade-offs. Implementing AI often requires substantial upfront investment in infrastructure, training, and governance mechanisms (Sani and Mandina, 2024). Moreover, disparities in access and institutional capacity risk reinforcing existing inequalities across public universities, particularly between well-funded and resource-constrained contexts (Yu et al., 2020). Ethical concerns, such as algorithmic bias, privacy issues, and the transparency of AI-driven decisions, also pose challenges for equitable adoption (Göçen et al., 2024). Accordingly, equity-by-design policies—procurement requirements for accessibility and privacy, routine bias audits for predictive systems, and targeted capacity-building grants for resource-constrained institutions—should accompany AI adoption so that documented savings advance digital-equity goals rather than exacerbate gaps.

**Table 2: Comparative Overview of Empirical Studies on AI and Cost Reduction in Public Higher Education**

| # | Authors | AI method / system | Cost Dimension | Outcome |
|---|---------|--------------------|----------------|---------|
| 1 | Vinnik & Scholl (2005) | UNICAP decision-support for capacity planning | Avoids costly misallocations; better room/staff utilization | Flagged over-capacity; enabled feasible admissions-target adjustment |
| 2 | Biazus et al. (2019) | Process mining within BPM | Shorter processing times; reduced staff workload; fewer delays | Identified bottlenecks/rework; guided workflow redesign implying lower admin spend |
| 3 | Cung et al. (2019) | ALEKS intelligent tutoring system (online vs blended) | Lower cost per student and per passing student | $53 online vs $124 blended; $142 per pass online vs $186 blended (with lower online outcomes) |
| 4 | Kumar (2019) | DEA + Kohonen self-organizing maps | Reduce slack; improve productivity without adding inputs | 14 institutes underutilized (mean DEA≈0.68 vs sample mean≈0.91); strategic levers identified |
| 5 | Casino et al. (2020) | Hybrid ray-launching + collaborative filtering for wireless modeling | Cuts computation/storage; cheaper planning for dense campus IoT | >40% data reduction; >30% DB pattern reduction; >8% accuracy gain |
| 6 | Aulck et al. (2020) | Enrollment prediction (ML) + genetic-algorithm scholarship optimization | Revenue uplift; reduces ad-hoc consulting/policy churn | Actual yield 14.8% vs 12% baseline (+23.3%); projected +15.8% over baseline |
| 7 | Phan et al. (2022) | Gradient boosting (AUC≈0.96) + multi-objective local-search optimization | Broaden access within budget; identify feasible aid policies | +31.9–32.3% expected enrollment; ~105–112% accessibility with ≤7% more aid |
| 8 | Wang (2023) | Multimodal engagement detection | Triage instructor time toward disengaging learners | Improved classification enabling targeted support |
| 9 | Abonizio et al. (2023) | CoronaAI WhatsApp chatbot (NLP/Dialogflow) | Handles routine queries at scale; reduces staff time; resilience during spikes | ~7,000 interactions/11 months; maintained service continuity with automated responses |
| 10 | Abdul-Rahman et al. (2023a) | Composite Resilience Index from social media + Delphi + AHP | Resource-efficient alternative to household surveys; cheaper recurrent monitoring | CRI built from 935,822 tweets; ≥70% weight on physical/economic/social/cultural criteria |

| # | Authors | AI method / system | Cost Dimension | Outcome |
|---|---------|--------------------|----------------|---------|
| 11 | Abdul-Rahman et al. (2023b) | Twitter mining + LDA topic modeling + VADER sentiment | Automates initial scoping vs. in-person consultations; faster/cheaper iteration | Ranked local challenges across six towns; "simple, fast, cheap" evidence pipeline |
| 12 | Rios-Esparza & Segura-Pérez (2023) | Hexagonal tessellation + simulation for placements | Cuts travel & admin labor; better matching | −585 km travel; +26% schedule match; +37% skills match; time cut 5 days → 90 min |
| 13 | Moreira & Freitas (2024) | Constraint-based timetabling with Google OR-Tools CP-SAT | Less manual rework; denser/stable room use; lower admin burden | Fewer constraint violations; improved utilization on real institutional data |
| 14 | Sani & Mandina (2024) | AI-supported digital HRM (screening, leave/payroll, routing) | Cuts manual handling/rework; cycle-time reductions; sustain service under tight budgets | Synthesizes barriers/enablers; practice guidance for time savings in core HR tasks |
| 15 | Göçen et al. (2024) | Sentiment & topic modeling of higher-ed tweets | Always-on "policy radar" to pre-empt costly issues | Finance/affordability themes surfaced; overall HE sentiment positive, AI mentions more negative |
| 16 | Tan & Dabbagh (2024) | IoT teaching platform (Kruskal MST + Gaussian classifier) | Automates orchestration; supports content reuse; lower marginal delivery cost | Experimental group showed higher understanding/engagement (vs control) |
| 17 | Wang & Zhang (2024) | VR edge-cloud classroom; DIBR-based collaborative transmission | Edge offloading reduces bandwidth/cloud costs; shared devices | VR group significantly higher across Bloom's levels (N=50 vs 50) |
| 18 | Nai et al. (2025) | ML forecasting on WB indicators (RF best) | Better budgeting/staffing; avoid overspend from misforecasted demand | RF: MAE=3.14, RMSE=3.91, $R^2$=0.92; sharper projections for dropout/support |
| 19 | Song et al. (2025) | AI-assisted thematic coding + semantic network analysis of e-texts | E-texts cut textbook/logistics costs; design preserves affordability | 322 responses: accessibility/portability valued; affordability emphasized by non-traditional students |
| 20 | Ugwu et al. (2025) | Adaptive, AI-based immersive self-directed English training | Substitutes instructor labor; lower unit cost at scale with fixed "contact hours" | AI group outperformed workshops at posttest & 3-month follow-up; large effects |
| 21 | Jaboob et al. (2025) | Generative-AI study tools | "Do more with the same" tutoring/contact hours | Positive paths to achievement; behavior partially mediates |

Despite the structured methodology, this study has several limitations. First, the evidence base is still emerging and relatively fragmented, with only 21 empirical studies meeting the inclusion criteria. Second, the review was limited to English-language, peer-reviewed publications indexed in Scopus and IEEE Xplore, raising the possibility of language and database bias. Third, the rapidly evolving nature of AI means that relevant studies may have been missed due to changes in terminology or publication lag. Finally, many of the included studies relied on small-scale or context-specific implementations, which may limit the generalizability of findings across diverse public higher education systems.

Future research should prioritize longitudinal and large-scale evaluations to assess the sustained financial and pedagogical effects of AI adoption in higher education. Comparative studies across regions and institutional types would provide valuable insights into how contextual factors shape efficiency gains and risks. Further investigation into governance, ethics, and equity is essential to ensure that AI does not exacerbate digital divides but instead contributes to inclusive and sustainable models of higher education (Abdul-Rahman et al., 2023; Song et al., 2025). Strengthening this evidence base will enable policymakers, administrators, and educators to make informed decisions about integrating AI for both economic efficiency and equitable access.

## 5   Conclusions

This scoping review demonstrates that AI holds significant potential to enhance economic efficiency and affordability in public higher education by streamlining administrative processes, optimizing resource allocation, supporting personalized learning, and strengthening predictive analytics for student success. At the same time, the evidence reveals important limitations and challenges, including substantial implementation costs, uneven institutional capacity, and the risk of exacerbating existing inequalities and digital divides. Collectively, these findings highlight both the opportunities and constraints of AI-driven cost reduction strategies. For policymakers, university administrators, and educators, the results underscore the need for careful planning, equity-oriented implementation, and ongoing evaluation to ensure that efficiency gains translate into sustainable and inclusive outcomes. Future research should expand empirical evidence across diverse contexts, explore long-term financial implications, and address governance and ethical considerations that shape the role of AI in higher education systems.

## References

Abdul-Rahman, M., Alade, W., & Anwer, S. (2023). A Composite Resilience Index (CRI) for Developing Resilience and Sustainability in University Towns. *Sustainability*, *15*(4), 3057. https://doi.org/10.3390/su15043057

Abdul-Rahman, M., Adegoriola, M. I., McWilson, W. K., Soyinka, O., & Adenle, Y. A. (2023b). Novel use of social media big data and artificial intelligence for community resilience assessment (CRA) in university towns. *Sustainability*, *15*(2), 1295. https://doi.org/10.3390/su15021295

Abonizio, H. Q., da Costa Barbon, A. P. A., Rodrigues, R., Santos, M., Martínez-Vizcaíno, V., Mesas, A. E., & Junior, S. B. (2023). How people interact with a chatbot against disinformation and fake nsonews in COVID-19 in Brazil: The CoronaAI case. *International Journal of Medical Informatics*, *177*, 105134.

https://doi.org/10.1016/j.ijmedinf.2023.105134

Al-Zahrani, A. M., & Alasmari, T. M. (2024). Exploring the impact of artificial intelligence on higher education: The dynamics of ethical, social, and educational implications. *Humanities and Social Sciences Communications*, *11*(1), 1-12. https://doi.org/10.1057/s41599-024-03432-4

Aulck, L., Nambi, D., & West, J. (2020). Increasing Enrollment by Optimizing Scholarship Allocations Using Machine Learning and Genetic Algorithms. In *Proceedings of the 13th International Conference on Educational Data Mining (EDM 2020)*. http://files.eric.ed.gov/fulltext/ED608000.pdf

Biazus, M., dos Santos, C. H., Takeda, L. N., de Oliveira, J. P. M., Fantinato, M., Mendling, J., & Thom, L. H. (2019). Software resource recommendation for process execution based on the organization's profile. In *International Conference on Database and Expert Systems Applications* (pp. 118-128). Springer, Cham. https://doi.org/10.1007/978-3-030-27618-8_9

Casino, F., Lopez-Iturri, P., Aguirre, E., Azpilicueta, L., Falcone, F., & Solanas, A. (2020). Enhanced wireless channel estimation through parametric optimization of hybrid ray launching-collaborative filtering technique. *IEEE Access*, *8*, 83070-83080. https://doi.org/10.1109/ACCESS.2020.2992033

Cung, B., Xu, D., Eichhorn, S., & Warschauer, M. (2019). Getting academically underprepared students ready through college developmental education: Does the course delivery format matter? *American Journal of Distance Education*, *33*(3), 178-194. https://doi.org/10.1080/08923647.2019.1582404

Göçen, A., Ibrahim, M. M., & Khan, A. U. I. (2024). Public attitudes toward higher education using sentiment analysis and topic modeling. *Discover Artificial Intelligence*, *4*(1), 83. https://doi.org/10.1007/s44163-024-00195-4

Goel, A. (2020). *AI-powered learning: making education accessible, affordable, and achievable*. arXiv preprint arXiv:2006.01908. https://doi.org/10.48550/arXiv.2006.01908

Haddaway, N. R., Page, M. J., Pritchard, C. C., & McGuinness, L. A. (2022). PRISMA2020: An R package and Shiny app for producing PRISMA 2020-compliant flow diagrams, with interactivity for optimised digital transparency and Open Synthesis. *Campbell systematic reviews*, *18*(2), e1230. https://doi.org/10.1002/cl2.1230

Jaboob, M., Hazaimeh, M., & Al-Ansi, A. M. (2025). Integration of generative AI techniques and applications in student behavior and cognitive achievement in Arab higher education. *International journal of human–computer interaction*, *41*(1), 353-366. https://doi.org/10.1080/10447318.2023.2300016

Kumar, S. (2019). Artificial intelligence divulges effective tactics of top management institutes of India. *Benchmarking: An International Journal*, *26*(7), 2188-2204. https://doi.org/10.1108/BIJ-08-2018-0251

Li, T. W., Karahalios, K., & Sundaram, H. (2021). " It's all about conversation" Challenges and Concerns of Faculty and Students in the Arts, Humanities, and the Social Sciences about Education at Scale. In *Proceedings of the ACM on Human-Computer Interaction*, *4* (CSCW3), 1-37. https://doi.org/10.1145/3432915

McConvey, K., Guha, S., & Kuzminykh, A. (2023, April). A human-centered review of algorithms in decision-making in higher education. In *Proceedings of the 2023 CHI Conference on Human Factors in Computing Systems* (pp. 1-15). https://doi.org/10.1145/3544548.3580658

Moreira, É. J. B., & De Freitas, S. A. A. (2024). A CP-SAT approach for academic resource timetabling in higher education institutions: A case study at a major public university. In *21st International Conference on Information Technology Based Higher Education and Training (ITHET)*, 1–8. https://doi.org/10.1109/ITHET61869.2024.10837617

Nai, S., Elbaghazaoui, B. E., Rifai, A., & Sadiq, A. (2025). Artificial intelligence predictive modeling for educational indicators using data profiling techniques. *IAES International Journal of Artificial Intelligence (IJ-AI), 14*(4), 3063–3073. https://doi.org/10.11591/ijai.v14.i4.pp3063-3073

Page, M. J., McKenzie, J. E., Bossuyt, P. M., Boutron, I., Hoffmann, T. C., Mulrow, C. D., ... & Moher, D. (2021). The PRISMA 2020 statement: an updated guideline for reporting systematic reviews. *bmj, 372*. https://doi.org/10.1136/bmj.n71

Phan, V., Wright, L., & Decent, B. (2022). Optimizing Financial Aid Allocation to Improve Access and Affordability to Higher Education. *Journal of Educational Data Mining, 14*(3), 26-51. http://files.eric.ed.gov/fulltext/EJ1373125.pdf

Rios-Esparza, G. A., & Segura-Pérez, E. (2023). A proposal of a simulation-optimization methodology for allocation of agencies with human resources on hexagonal tessellation. *Journal of applied research and technology, 21*(1), 106-122. https://doi.org/10.22201/icat.24486736e.2023.21.1.2183

Sani, S., & Mandina, S. P. (2024). Examining the Challenges of Adopting Modern Technologies in Public Sector Human Resource Management. In *Digital Transformation in Public Sector Human Resource Management* (pp. 60-89). IGI Global. https://doi.org/10.4018/979-8-3693-2889-7.ch004

Sohrabi, C., Franchi, T., Mathew, G., Kerwan, A., Nicola, M., Griffin, M., ... & Agha, R. (2021). PRISMA 2020 statement: What's new and the importance of reporting guidelines. *International Journal of Surgery, 88*, 105918. https://doi.org/10.1016/j.ijsu.2021.105918

Song, S., Kim, H., Simpson, L., & Rivas, J. (2025). Non-traditional and traditional business students' perceptions of e-texts: A computer-assisted analysis with implications for marketing education. *Marketing Education Review*, 1-17. https://doi.org/10.1080/10528008.2025.2545805

Tan, C., & Dabbagh, N. (2024). Construction and Application of an Interactive Teaching Platform for Aesthetic Education in Ecological Environments. *International Journal of Web-Based Learning and Teaching Technologies (IJWLTT), 19*(1), 1-16. https://doi.org/10.4018/IJWLTT.339185

Ugwu, N. F., Ochiaka, R. E., Asogwa, U. S., Igbinlade, A. S., Sanni, K. T., Onayinka, T. S., ... & Olubodun, O. A. (2025). Comparing the Efficacy of Artificial Intelligence Immersion and Human-Led Workshops for Enhancing Researchers' English Language Skills: A Randomized Control Trial. *Higher Learning Research Communications, 15*(1), 2. https://doi.org/10.18870/hlrc.v15i1.1530

Vinnik, S., & Scholl, M. H. (2005, March). UNICAP: Efficient decision support for academic resource and capacity management. In *International Conference on e-Government* (pp. 235-246). Berlin, Heidelberg: Springer Berlin Heidelberg. https://doi.org/10.1007/978-3-540-32257-3_22

Wang, Z. (2023). Higher education management work based on information fusion technology for the development of innovation ability of college students. *Applied Mathematics and Nonlinear Sciences, 9*(1), 2023. https://doi.org/10.2478/amns.2023.2.00775

Wang, S. & Zhang, S. (2024). Optimization of Higher Education Teaching Method System Based on Edge Intelligence. *Applied Mathematics and Nonlinear Sciences, 9*(1), 2024. https://doi.org/10.2478/amns-2024-3007

Yu, R., Li, Q., Fischer, C., Doroudi, S., & Xu, D. (2020). Towards Accurate and Fair Prediction of College Success: Evaluating Different Sources of Student Data. *In Proceedings of International Conference on Educational Data Mining*. https://files.eric.ed.gov/fulltext/ED608066.pdf